\documentclass[runningheads]{llncs}
\usepackage[T1]{fontenc}
\usepackage{graphicx}
\usepackage{hyperref}
\usepackage{color}

\usepackage{physics}
\usepackage{comment}
\usepackage{amsmath,amsfonts,amssymb}

\def\fcmp{\mathbin{\raise 0.6ex\hbox{\oalign{\hfil$\scriptscriptstyle      \mathrm{o}$\hfil\cr\hfil$\scriptscriptstyle\mathrm{9}$\hfil}}}}

\newcommand{\cat}[1]{\ensuremath{\mathbf{#1}}}
\newcommand{\id}{\ensuremath{\mathrm{id}}}
\newcommand{\hlam}[2]{\langle #1 ~|~ #2 \rangle}
\newcommand{\PiLang}{\ensuremath{\Pi}}
\newcommand{\fromto}{\leftrightarrow}
\newcommand{\defeq}{\mathbin{::=}}
\newcommand{\cm}{\mathit}
\newcommand{\of}{\mathbin{:}}

\newcommand{\hilbspace}{\ensuremath{A}}

\newcommand{\seqq}{\fcmp}
\newcommand{\pid}{\cm{id}}
\newcommand{\swapp}{\cm{swap}^{+}}
\newcommand{\assocp}{\cm{assocr}^+}
\newcommand{\associp}{\cm{assocl}^+}
\newcommand{\unitep}{\cm{unite}^+}

\newcommand{\unitepl}{\cm{unite}^{+}l}
\newcommand{\unitipl}{\cm{uniti}^{+}l}
\newcommand{\swapt}{\cm{swap}^{\times}}
\newcommand{\assoct}{\cm{assocr}^{\times}}
\newcommand{\associt}{\cm{assocl}^{\times}}
\newcommand{\unitet}{\cm{unite}^{\times}}

\newcommand{\unitetl}{\cm{unite}^{\times}l}
\newcommand{\unititl}{\cm{uniti}^{\times}l}
\newcommand{\dist}{\cm{dist}}
\newcommand{\factor}{\cm{factor}}
\newcommand{\absorbl}{\cm{absorbl}}
\newcommand{\factorzr}{\cm{factorzr}}
\newcommand{\inv}{\cm{inv}}

\newcommand{\ppp}{*\!\!*\!\!*}

\newcommand{\reducesto}{\rightarrow}
\newcommand{\reducesfrom}{\leftarrow}

\begin{document}
\title{Compositional Reversible Computation}
%
%
\author{Jacques Carette\inst{1}\orcidID{0000-0001-8993-9804} \and
Chris Heunen\inst{2}\orcidID{0000-0001-7393-2640} \and
Robin Kaarsgaard\inst{3}\orcidID{0000-0002-7672-799X} \and
Amr Sabry\inst{4}\orcidID{0000--0002-1025-7331}
}
\authorrunning{J. Carette et al.} 
%
\institute{McMaster University, Hamilton, Ontario, Canada \and
  University of Edinburgh, Edinburgh, UK \and
  University of Southern Denmark, Odense, Denmark \and
  Indiana University, Bloomington, Indiana, USA}

\maketitle              
\begin{abstract}
  Reversible computing is motivated by both pragmatic and foundational
  considerations arising from a variety of disciplines. We take a
  particular path through the development of reversible computation,
  emphasizing \emph{compositional} reversible
  computation.  We start from a historical perspective, by reviewing
  those approaches that developed reversible extensions of
  $\lambda$-calculi, Turing machines, and communicating process
  calculi. These approaches share a common challenge: computations
  made reversible in this way do not naturally compose locally.

  We then turn our attention to computational models that eschew the
  detour via existing irreversible models. Building on an original
  analysis by Landauer, the insights of Bennett, Fredkin, and Toffoli
  introduced a fresh approach to reversible computing in which
  reversibility is elevated to the status of the main design
  principle.  These initial models are expressed using low-level bit
  manipulations, however.

  Abstracting from the low-level of the
  Bennett-Fredkin-Toffoli models and pursuing more intrinsic, typed,
  and algebraic models, naturally leads to rig categories as the canonical
  model for compositional reversible programming. The categorical
  model reveals connections to type isomorphisms, symmetries,
  permutations, groups, and univalent universes. This, in turn, paves 
  the way for extensions to reversible programming based on monads and
  arrows. These extensions are shown to recover conventional
  irreversible programming, a variety of reversible computational
  effects, and more interestingly both pure (measurement-free) and
  measurement-based quantum programming.

\keywords{Rig Categories \and Information Effects \and Quantum Computing.} 
\end{abstract}
%
%
\section{Introduction}

In 1992, Baker proposed ``an abstract computer model and a programming
language--\textsf{$\Psi$-Lisp}--whose primitive operations are
injective and hence reversible''~\cite{10.5555/645648.664816}.  The
proposal was motivated by both software engineering and physics.

The software engineering perspective, building on earlier insights by
McCarthy~\cite{McC56} and Zelkowitz~\cite{10.1145/362342.362360},
recognizes that reversibility is a pervasive occurrence in a large
number of programming activities:
\begin{quote}
  The need to reverse a computation arises in many
  contexts--debugging, editor undoing, optimistic concurrency undoing,
  speculative computation undoing, trace scheduling, exception
  handling undoing, database recovery, optimistic discrete event
  simulations, subjunctive computing, etc. The need to analyze a
  reversed computation arises in the context of static
  analysis--liveness analysis, strictness analysis, type inference,
  etc. Traditional means for restoring a computation to a previous
  state involve checkpoints; checkpoints require time to copy, as well
  as space to store, the copied material. Traditional reverse abstract
  interpretation produces relatively poor information due to its
  inability to guess the previous values of assigned-to
  variables. 
\end{quote}

The more foundational physics perspective recognizes that a ``physics
revolution is brewing in computer science.'' This ``physics
revolution'' traces to developments started 20 years earlier beginning
with an analysis of logical (ir)reversibility and its connection to
physical (ir)reversibility by Landauer~\cite{5392446}. This initial
analysis demonstrated how an \emph{isolated} irreversible computation
can be embedded in a larger reversible one but failed to solve the
problem of composing such embeddings. The solution to this puzzle was
provided a decade later by Bennett~\cite{5391327}; it involved a
general idiom \textsf{compute-copy-uncompute} that proved crucial for
further developments. A few years later, Fredkin and
Toffoli~\cite{fredkin-conservative-1982,10.1007/3-540-10003-2104}
finally designed a foundational model of composable reversible
computations based purely on reversible primitives.

In our survey of part of the landscape of reversible computation, we
start by reviewing the research initiatives whose goal is to develop a
reversible programming language starting from existing (irreversible)
languages. We will then consider the more foundational idea of taking
reversibility as the main primitive notion, formalizing it, and
extending it in principled ways to realize the full potential of the
``physics revolution in computer science.'' In more detail,
Sec.~\ref{sec:global} reviews the early historical proposals for
reversible computing characterized by using global history
mechanisms. Sec.~\ref{sec:local} discusses one of the crucial ideas
necessary for compositional reversible computing: the
compute-copy-uncompute paradigm. Sec.~\ref{sec:rig} exploits the power
of categorical semantics to naturally express compositional reversible
computing. Sec.~\ref{sec:effects} discusses general classes of
reversible computational effects concluding with the ``fundamental
theorem of reversible computation.'' Sec.~\ref{sec:quantum} shows that
the categorical models of classical reversible computing with computational effects extend to quantum computing. We conclude
with an assessment of the broad impacts of ``reversibility'' on the
discipline of computer science.

\section{Reversibility from Global Histories} 
\label{sec:global}

The most familiar sequential models of computation are the Turing
machine and the $\lambda$-calculus. Both were proposed in the
1930s~\cite{15897363-af72-3dac-82e6-fde144ad66c0,doi.org/10.1112/plms/s2-42.1.230}. In
the concurrent world, we have the influential models of
\emph{Communicating Sequential Processes} (CSP), the \emph{Calculus of
  Communicating Systems} (CCS), and the
$\pi$-calculus~\cite{10.1145/359576.359585,DBLP:books/sp/Milner80,Milner_1992}.

The methods used to derive a reversible variant of these models of
computation are
similar. In each case, additional constructs are added to record the
information necessary for reversibility. In what follows, we discuss
how to do this for the reversible extension of the
$\lambda$-calculus~\cite{doi:10.1137/S0097539703432165} and the
reversible extensions of CCS called \emph{Reversible CCS} (RCCS) and
\emph{CCS with keys}
(CCSK)~\cite{10.1007/978-3-540-28644-8_19,10.1007/11690634_17}.

The operational semantics of both sequential and concurrent
programming languages is often specified using local reductions, e.g.,
$A \reducesto B$. In a deterministic language $A$ uniquely determines
$B$ but if the language is not reversible, the converse is not
true. In other words, it is possible for have instances of reductions
where both $A_1 \reducesto B$ and $A_2 \reducesto B$.

A straightforward way to ensure each reduction is reversible is to
record additional information to disambiguate the lefthand sides. In
the simplest case, we introduce a history mechanism $H$ where we
record the entire term on the lefthand side, i.e., the reductions
above become:
\[\begin{array}{rcll}
\hlam{H}{A_1} &\reducesto& \hlam{H,A_1}{B} \\
\hlam{H}{A_2} &\reducesto& \hlam{H,A_2}{B}
\end{array}\]
An adequate history mechanism disambiguates which path the 
computation took to get to $B$, so that we now have enough
information to reverse the reductions:
\[\begin{array}{rcll}
\hlam{H}{A_1} &\reducesfrom_H& \hlam{H,A_1}{B} \\
\hlam{H}{A_2} &\reducesfrom_H& \hlam{H,A_2}{B}
\end{array}\]

This simple scheme can be optimized in many ways to manage the history
more efficiently. However, a fundamental limitation of this approach
is that it fails to be \emph{compositional}:
Consider a term that includes both $A_1$ and $A_2$ as subterms
and where $A_1$ should make a forward transition and $A_2$ make a
backwards transition. Both sub-reductions require incompatible actions
on the global history mechanism and direct composition is not
possible. As the analysis of this problem in the context of CCSK
shows~\cite{AubertRCC23}, the best solution is to take reversibility as a
basic building block, instead of a property to be achieved by
extensions to an irreversible language.

\section{Reversibility from Local Histories} 
\label{sec:local}


In keeping with the approach above, Landauer observed that any Turing
machine can be altered to operate reversibly by adding a dedicated
\emph{history tape} to it, recording each computational action on this
tape as it occurs.  However, Landauer also observed that, from a
thermodynamic point of view, this approach is fundamentally
unsatisfactory in that it merely \emph{delays} rather than
\emph{avoids} the thermodynamic cost associated with the erasure of
unwanted information~\cite{5392446}: to be able to reuse the tape,
its contents must first be erased. To Bennett, this meant that the
usefulness of a reversible computer hinged on the ability to avoid
this problem, leaving behind ``only the desired output and the
originally furnished input'' when it halts~\cite{5391327} (remarking
that the preservation of the input is necessary to realize computable
functions which happen to not be injective). This was likely the first
instance of \emph{reversibility as compositionality}, and has since
been rediscovered numerous times in the context of circuits,
programming languages, and categorical semantics.

It is amusing to note that ``undo/redo'' functionality in modern
user-interfaces use either the \emph{Command Pattern} or the
\emph{Memento Pattern}, which both amount to the non-composable
Landauer encoding. 

\subsection{Bennett's Trick}

The key insight behind Bennett's trick is that the use of
\emph{uncomputation} (i.e., inverse interpretation of a \emph{reversible}
Turing machine) can reduce the dependence on a computation history to
the preservation of the input. This is done by proceeding in three
stages (see Figure~\ref{fig:bennetttrick}): \emph{compute} executes
the Turing machine to obtain the output and its history, \emph{copy}
copies the output onto a dedicated output tape (assumed to be empty),
and \emph{uncompute} executes the Turing machine in reverse to reduce
the output and history to the original input.

If we think of the computation history more generally as the garbage
that is inevitably produced during computation (i.e., temporary
storage needed during computation that can safely be discarded
afterwards), Bennett's trick gives a reversible way of managing
garbage without having to erase it outright. This allows procedures
that use the same pool of temporary storage to be composed without
incident, as they can all safely assume the store to be empty when
needed. This technique is used to manage memory in, e.g., reversible
programming languages~\cite{10.1007/978-3-642-19861-8_9} and
reversible circuits~\cite{10.1007/978-3-319-20860-2_13}.

\begin{figure}[t]
\centering
\begin{tabular}{p{5cm} l l l}
  \textbf{Stage} & \textbf{Tape 1} & \textbf{Tape 2} & \textbf{Tape 3} \\
  \hline
  Initial configuration & Input & -- & -- \\
  Compute & Output & History & -- \\
  Copy & Output & History & Output \\
  Uncompute & Input & -- & Output \\
\end{tabular}
\caption{Bennett's construction of a standard reversible 3-tape
  Turing machine, starting from an arbitrary Turing machine
  instrumented to record its history on a dedicated tape.}
\label{fig:bennetttrick}
\end{figure}

\begin{figure}[t]
\centering
\begin{tabular}{p{5cm} l l}
  \textbf{Stage} & \textbf{Tape 1} & \textbf{Tape 2} \\
  \hline
  Initial configuration & Input & -- \\
  Copy & Input & Input \\
  Compute & Output & Input \\
  Relabel & Input & Output \\
\end{tabular}
\caption{An overly simple 2-tape Turing machine that ``looks''
  like it operationally does the same as Bennett's construction.}
\label{fig:toosimple}
\end{figure}

Na\"{\i}vely, one might suppose that a simpler approach (see
Figure~\ref{fig:toosimple}) might work too. The problem is that this
only works when the Turing machine is reversible to begin with!
Furthermore, what is ``Relabel''? Is that even an available operation
on Turing machines? We could think of replacing Relabel with some kind
of Swap operation, but since there is no guarantee that Input and
Output are the same length, this operation is not reversible either
without further assumptions. Another way to look at it: if one were
working in a dependently typed language, Bennett's construction would
require a proof that the history information is sufficient to actually
drive the given Turing machine \emph{deterministically} backwards.


\subsection{Reversibility as a Local Phenomenon}

While it seems clear that maintaining a history during
computation is a general method for guaranteeing that the resulting
Turing machine is reversible, it doesn't actually answer what it means
for a Turing machine to be reversible in the first place.

Lecerf~\cite{lecerf1963machines} answered this by defining a
reversible Turing machine to be one where at each computational state,
there is at most one \emph{next} state and at most one \emph{previous}
state. We can express this more precisely using the judgement $\sigma
\vdash c \downarrow \sigma'$, taken to mean that executing the command
$c$ while the machine is in state $\sigma$ leads the machine to
transition to the state $\sigma'$. The unicity of the next and
previous states become the statements (see, e.g.,
\cite{GLUCK2023113429}) that for all commands~$c$ and origin states
$\sigma$, there is \emph{at most} one next state $\sigma'$ such that
$\sigma \vdash c \downarrow \sigma'$ (forward determinism); and for
all commands $c$ and states $\sigma'$, there is \emph{at most} one
origin state $\sigma$ such that $\sigma \vdash c \downarrow \sigma'$
(backward determinism).

This establishes reversibility as a local phenomenon linked directly
to compositionality: it is not enough to compute an injective function
(a global property) to be reversible, it must also be done by taking
only invertible steps along the way. Indeed, a defining consequence of
this very strong conception of reversibility (amusingly dubbed the
``Copenhagen interpretation'' of reversible computation by
Yokoyama~\cite{GLUCK2023113429,HBA}) is the property of \emph{local
invertibility}, allowing a reversible machine (or program) to be
inverted by recursive descent over the
syntax~\cite{10.1007/978-3-540-24754-8_21}. This idea was taken to its
logical conclusion in the (explicitly compositional) denotational
account of reversibility~\cite{RKthesis}, where it was argued that a
program should be considered to be reversible just in case it can be
constructed by combining only invertible parts in ways that preserve
invertibility. A reasonable place to take such denotational semantics
is in categories of invertible maps, such as inverse
categories~\cite{kastl:inverse,kaarsgaardglueck:rfcl} and
groupoids~\cite{CARETTE202215,10.1007/978-3-662-49498-1,abramsky:reversible}.

\section{Rig Groupoids}
\label{sec:rig}

%

Category theory deals with abstractions in a uniform and systematic
way, and is widely used to provide compositional programming
semantics. We briefly discuss the types of categories that are useful
in reversible programming: dagger categories and rig categories. 

\subsection{Dagger Categories and Groupoids}

A morphism $f \colon A \to B$ is \emph{invertible}, or an
\emph{isomorphism}, when there exists a morphism
$f^{-1} \colon B \to A$ such that $f^{-1} \circ f = \id_A$ and
$f \circ f^{-1} = \id_B$. This inverse $f^{-1}$ is necessarily
unique. A category where every morphism is invertible is called a
\emph{groupoid}. At first sight, groupoids form the perfect semantics
for reversible computing. But every step in a computation being
reversible is slightly less restrictive than it being invertible. For
each step $f \colon A \to B$, there must still be a way to `undo' it,
given by $f^\dag \colon B \to A$. This should also still respect
composition, in that $(g \circ f)^\dag = f^\dag \circ g^\dag$ and
$\id_A^\dag = \id_A$. Moreover, a `cancelled undo' should not change
anything: $f^{\dag\dag}=f$. Therefore every morphism $f$ has a partner
$f^\dag$.  A category equipped with such a choice of partners is
called a \emph{dagger category}.

A groupoid is an example of a dagger category, where every morphism is
\emph{unitary}, that is, $f^\dag = f^{-1}$. Think, for example, of the category
$\cat{FinBij}$ with finite sets for objects and bijections for morphisms. But
not every dagger category is a groupoid. For example, the dagger category
$\cat{PInj}$ has sets as objects, and partial injections as morphisms. Here,
the dagger satisfies $f \circ f^\dag \circ f = f$, but not necessarily $f^\dag
\circ f = \id$ because $f$ may only be partially defined. In a sense, the
dagger category $\cat{PInj}$ is the universal model for reversible partial
computation~\cite{kastl:inverse,heunen:ltwo}.

When a category has a dagger, it makes sense to demand that every
other structure on the category respects the dagger, and we will do
so. The theory of dagger categories is similar to the theory of
categories in some ways, but very different in
others~\cite{heunenkarvonen:daggermonads}.

\subsection{Monoidal Categories and Rig Categories}

Programming becomes easier when less encoding is necessary, i.e. when
there are more first-class primitives.  For example, it is handy to
have type combinators like sums and products. Semantically, this is
modeled by considering not mere categories, but monoidal ones. A
\emph{monoidal category} is a category equipped with a type combinator
that turns two objects $A$ and $B$ into an object $A \otimes B$, and a
term combinator that turns two morphisms $f \colon A \to B$ and
$f' \colon A' \to B'$ into a morphism
$f \otimes f' \colon A \otimes A' \to B \otimes B'$. This has to
respect composition and identities. Moreover, there has to be an
object $I$ that acts as a unit for~$\otimes$, and isomorphisms
$\alpha \colon A \otimes (B \otimes C) \to (A \otimes B) \otimes C$
and $\lambda \colon I \otimes A \to A$ and
$\rho \colon A \otimes I \to A$. In a \emph{symmetric monoidal
  category}, there are additionally isomorphisms
$\sigma \colon A \otimes B \to B \otimes A$. All these isomorphisms
have to respect composition and satisfy certain coherence conditions,
see~\cite{maclane:associativity} or~\cite[Chapter
1]{heunenvicary:cqt}. We speak of a \emph{(symmetric) monoidal dagger
  category} when the coherence isomorphisms are unitary.  Intuitively,
$g \circ f$ models sequential composition, and $f \otimes g$ models
parallel composition. For example, $\cat{FinBij}$ and $\cat{PInj}$ are
symmetric monoidal dagger categories under cartesian product.

A \emph{rig category} is monoidal in two ways in a distributive
fashion. More precisely, it has two monoidal structures $\oplus$ and
$\otimes$, such that $\oplus$ is symmetric monoidal but $\otimes$ not
necessarily symmetric, and there are isomorphisms
$\delta_L \colon A \otimes (B \oplus C) \to (A \otimes B) \oplus (A
\otimes C)$ and $\delta_0 \colon A \otimes 0 \to 0$. These
isomorphisms again have to respect composition and certain coherence
conditions~\cite{laplaza:distributivity}. For example, $\cat{FinBij}$
and $\cat{PInj}$ are not only monoidal under cartesian product, but
also under disjoint union, and the appropriate distributivity
holds.  Intuitively, given $f : A \rightarrow B$ and $g : C \rightarrow D$,
$f \oplus g : A \oplus C \rightarrow B \oplus D$ models a choice between 
$f$ and $g$, predicated on whether it gets an $A$ or a $B$ as choice of
input.

\subsection{The Canonical Term Model $\Pi$} 

\begin{figure}[t]
\begin{align*}
  b & \defeq 0 \mid 1 \mid b+b \mid b \times b & \text{(value types)}
  \\ t & \defeq b \fromto b & \text{(combinator types)} \\ i & \defeq
  \pid \mid \swapp \mid \assocp \mid \associp \mid \unitepl \mid
  \unitipl & \text{(isomorphisms)} \\ & \mid \swapt \mid \assoct \mid
  \associt \mid \unitetl \mid \unititl \\ & \mid \dist \mid \factor
  \mid \absorbl \mid \factorzr \\ c & \defeq i \mid c \seqq c \mid c +
  c \mid c \times c \mid \inv~c & \text{(combinators)}
\end{align*}
\caption{\label{fig:pi}$\PiLang$ syntax}
\end{figure}

\begin{figure}[t]
\begin{equation*}
  \begin{array}{rcrclcl}
    \pid & \of & b &\fromto& b & \of & \pid \\
    \swapp & \of & b_1 + b_2 &\fromto& b_2 + b_1 & \of & \swapp \\
    \assocp & \of & (b_1 + b_2) + b_3 &\fromto& b_1 + (b_2 + b_3) & 
    \of & \associp \\
    \unitepl & \of & 0 + b &\fromto& b & \of & \unitipl \\
    \swapt & \of & b_1 \times b_2 &\fromto& b_2 \times b_1 & \of &
    \swapt \\
    \assoct & \of & (b_1 \times b_2) \times b_3 &\fromto& b_1 \times
    (b_2 \times b_3) & \of & \associt \\
    \unitetl & \of & 1 \times b &\fromto& b & \of & \unititl \\
    \dist & \of & (b_1 + b_2) \times b_3 &\fromto& (b_1 \times b_3)
    + (b_2 \times b_3) & \of & \factor \\
    \absorbl & \of & b \times 0 &\fromto& 0 & \of & \factorzr
  \end{array}
\end{equation*}
\begin{equation*}
  \frac{c_1 \of b_1 \fromto b_2 \quad c_2 \of b_2 \fromto b_3}{c_1 \seqq c_2 
    \of b_1 \fromto b_3}
  \qquad\qquad\qquad
  \frac{c \of b_1 \fromto b_2}{\inv~c \of b_2 \fromto b_1}
\end{equation*}
\begin{equation*}
  \frac{c_1 \of b_1 \fromto b_3 \quad c_2 \of b_2 \fromto b_4}{c_1 + c_2 
  \of b_1 + b_2 \fromto b_3 + b_4} \qquad
  \frac{c_1 \of b_1 \fromto b_3 \quad c_2 \of b_2 \fromto b_4}{c_1 \times c_2 
  \of b_1 \times b_2 \fromto b_3 \times b_4}
\end{equation*}
\caption{\label{fig:pitypes}Types for $\PiLang$ combinators}
\end{figure}

Given a rig groupoid, we may think of the objects as types, and the
morphisms as terms~\cite{leinster:categorytheory}. The syntax of the
language~$\PiLang$ in Fig.~\ref{fig:pi} captures this idea.  Type
expressions $b$ are built from the empty type (0), the unit type (1),
the sum type ($+$), and the product type ($\times$). A type
isomorphism $c : b_1 \fromto b_2$ models a reversible function that
permutes the values in $b_1$ and $b_2$. These type isomorphisms are
built from the primitive identities $i$ and their compositions. These
isomorphisms correspond exactly to the laws of a \emph{rig}
operationalised into invertible
transformations~\cite{10.1007/978-3-662-49498-1,CARETTE202215} which
have the types in Fig.~\ref{fig:pitypes}. Each line in the top part of
the figure has the pattern $c_1 : b_1 \fromto b_2 : c_2$ where $c_1$
and $c_2$ are self-duals; $c_1$ has type $b_1 \fromto b_2$ and~$c_2$
has type $b_2 \fromto b_1$.

To recap, the ``groupoid'' structure arises when we want all terms of
our programming language to be typed, reversible and composable. The
``rig'' part, is a pun on ``ring'' where the removal of the ``n''
indicates that we do not have \emph{negatives}.  The multiplicative
structure is used to model \emph{parallel} composition, and the
additive structure is a form of \emph{branching} composition. All this
structure is essentially forced on us once we assume that we want to
work over the (weak) semiring of finite types with products and
coproducts. We remark that in the categorical setting of $\PiLang$,
there are no issues in having terms like $(\assoct \times \factor)$ of
the form $(c_1 \times c_2)$ with $c_1$ is an isomorphism in the
forward direction and $c_2$ is an isomorphism in the reverse
direction. Composition is natural!

\subsection{Finite Sets and Permutations} 

It is folklore that the groupoid of finite sets and permutations is
the free symmetric rig groupoid on zero
generators~\cite{baez-dolan,10.1007/BFb0063106,laplaza:distributivity}.
Given that the syntax of $\PiLang$ is presented by the free symmetric
rig groupoid, given by finite sets and permutations, the folklore
result can be formally established ~\cite{10.1145/3498667}. The formal
connection provides an equational theory for $\PiLang$ that exactly
includes all the necessary equations to decide equivalence of
$\PiLang$ programs. 

\subsection{Curry-Howard}

Reversible computation also brings new light to the Curry-Howard
correspondence. The original correspondence, between type theory and
logic, formally focuses on \emph{equi-inhabitation}.  This is because
while logically $A$ and $A \land A$ (as well as $A$ and $A \lor A$)
are logically equivalent, clearly, as types, $A$ and $A \cross A$
(similarly, $A$ and $A + A$) are not equivalent.  They are, however,
equi-inhabited, i.e we can show that $a : A$ if and only if
$b : A \cross A$, but the witnessing functions are not inverses. We
can thus say that the Curry-Howard correspondence focuses on logically
equivalent types (often denoted $A \Leftrightarrow B$).

In $\PiLang$, we replace logical equivalence by equivalence $A \simeq B$.
And what used to be a correspondence between classical type theory
(involving types, functions and logical equivalence) and logic, transforms 
into a correspondence between reversible type theory (involving types,
reversible functions and equivalences) and \emph{algebra}, in this case
rigs and their categorified cousins, rig categories. This picture emerged
in the first looks at $\PiLang$~\cite{10.1007/978-3-662-49498-1,CARETTE202215}
and was shown to be \emph{complete} more recently~\cite{10.1145/3498667}.
In other words, $\PiLang$ is the inevitable programming language that
arises from universal reversible computing being semantically
about $\cat{FinBij}$.

\section{Reversible and Irreversible Effects}
\label{sec:effects}

Expressing reversibility in a categorical setting enables the integration
of additional constructs using universal categorical constructions such
as monads and arrows. 

\subsection{Frobenius Monads and Reversible Arrows}

So far, we have modeled computations as morphisms in a category. But
often it makes sense to separate out specific aspects of computation,
distinguishing between \emph{pure} computations, that only concern
themselves with computing values, and \emph{effectful} computations,
that can additionally have side effects, such as interacting with
their environment through measurement.

A \emph{monad} $T$ is a way to encapsulate computational side effects
in a modular way. If $A$ is an type, then $T(A)$ is the type of $A$
with possible side effects. For example, for the \emph{maybe} monad
$T(A)=A+1$, a term of type $T(A)$ is either a term of type $A$ or the
unique term of type $1$, which may be thought of as an exception
having occurred.

Regarding morphisms $A \to B$ as pure computations, computations that
can have side effects governed by $T$ are then morphisms $A \to
T(B)$. For this to make sense, we need three ingredients, which are
what makes $T$ into a monad: first, a way to consider a pure morphism
$A \to B$ as an effectful one $A \to T(B)$; second, to lift a pure
morphism $A \to B$ to the effectful setting $T(A) \to T(B)$; and
third, a way to sequence effectful computations $f \colon A \to T(B)$
and $g \colon B \to T(C)$ into $f \ggg g \colon A \to
T(C)$. 
The resulting category of effectful computations is called the
\emph{Kleisli category} of the monad $T$~\cite{moggi}.

What about the reversible setting? If the category of pure computations has a dagger, when does the category of effectful computations have a dagger that extends the reversal of pure computations? It turns out that this can be captured neatly in terms of the monad alone. The Kleisli category has a dagger if and only if the monad is a \emph{dagger Frobenius monad}~\cite{heunenkarvonen:daggermonads}, meaning that
\[
  T(f)^\dag = T(f^\dag) \qquad \text{ and }\qquad
  T(\mu_A) \circ \mu_{T(A)}^\dag = \mu_{T(A)} \circ T(\mu_A^\dag)\text,
\]
where $\mu_A \colon T(T(A)) \to T(A)$ is the sequencing of the identity $T(A) \to T(A)$, regarded as an effectful computation from $T(A)$ to $A$, with itself.

More generally, we can talk about \emph{arrows} instead of
monads. These still allow a sequential composition of effectful
computations~\cite{hughes:arrows,jacobsheunenhasuo:arrows}, and still
extend to the reversible
setting~\cite{heunenkaarsgaardkarvonen:arrows}.

\subsection{The Fundamental Theorem of Reversible Computation}
\label{sec:fundamental}

In this section, we define and further discuss the fundamental theorem
of reversible computing in terms of \emph{universal properties}. These
are categorical properties that characterize the result of some
construction in terms of its behavior only and not in terms of the
particular construction itself. For example, singleton sets $1$ are
characterized by the fact that there is a unique function $A \to 1$
for any set $A$; notice that the property only speaks about morphisms
into 1, and never about how 1 is built of a single element. We say
that 1 is a \emph{terminal object} in the category of sets and
functions. In the opposite direction, the empty set is an
\emph{initial object}, meaning that there is a unique function
$0 \to A$ for any set $A$. Similarly, the cartesian product
$A \times B$ of sets can be characterized universally as a categorical
\emph{product} of $A$ and $B$, and the disjoin union $A+B$ as a
\emph{coproduct}: these are the universal objects equipped with
projections $A \leftarrow A \times B \to B$ respectively injections
$A \rightarrow A+B \leftarrow B$~\cite{leinster:categorytheory}.

The fundamental theorem of reversible computing (originally proved by
Toffoli for functions over finite collections of boolean
variables~\cite{10.1007/3-540-10003-2104}) can now be phrased
categorically as follows~\cite{heunenkaarsgaard:qie}.

We first recall the
\emph{LR-construction}~\cite{andresmartinezheunenkaarsgaard:universal}
which turns a rig category into another category where $I$ is terminal
and $0$ is initial. In more detail, morphisms $A \to B$ in the new
category $\cat{LR}[\cat{C}]$ are morphisms
$A \oplus H \to B \otimes G$ in the old category $\cat{C}$, which we
identify if they behave similarly on the `heap' $H$ and `garbage' $G$.
The fundamental theorem of reversible computing follows by noting that
there is an inclusion from the category $\cat{Bij}$ of sets and
bijections to the category $\cat{Set}$ of sets and all
functions. Then, by the universal property of the LR-construction,
this inclusion factors through a functor
$\cat{LR}[\cat{Bij}] \to \cat{Set}$. Any function in $\cat{Set}$ is in
the image of this functor. In other words, any function
$f \colon A \to B$ is of the form
\[
  A \stackrel{i}{\longrightarrow} A + H \stackrel{\simeq}{\longrightarrow} B \times G \stackrel{\pi}{\longrightarrow} B\text,
\]
where $i$ is a coproduct injection, $\pi$ is a product projection, $G$
is the garbage, and the function in the middle is a
bijection.\footnote{As noted earlier, Toffoli only proved this for
  finite sets. We thank Tom Leinster for the following neat proof for
  infinite sets. In a category with products, every morphism
  $f \colon A \to B$ factors as a split monic
  $(1,f) \colon A \to A \times B$ followed by a product projection
  $A \times B \to B$. But in $\cat{Set}$, split monics are exactly the
  same as injections, which are coproduct injections up to an
  isomorphism.  }

\section{Quantum Effects} 
\label{sec:quantum}

It turns out that the standard model of quantum computing is a dagger
rig category. It is therefore natural to investigate the
classical-quantum connection(s) by investigating the corresponding
instances of rig categories.

\subsection{The Hilbert Space Model}
\label{sub:quantise}

Quantum computing with pure states is a specific kind of reversible
computing~\cite{yanofskymannucci:quantumcomputing,nielsenchuang:quantum}.
A quantum system is modeled by a finite-dimensional Hilbert
space~\hilbspace. For example, \emph{qubits} are modeled by
$\mathbb{C}^2$. The category giving semantics to finite-dimensional
pure state quantum theory is therefore $\cat{FHilb}$, whose objects
are finite-dimensional Hilbert spaces, and whose morphisms are linear
maps. Categorical semantics for pure state quantum computing is the
groupoid $\cat{Unitary}$ of finite-dimensional Hilbert spaces as
objects with unitaries as morphisms. Both are rig categories under
direct sum $\oplus$ and tensor product $\otimes$.

The pure \emph{states} of a quantum system modeled by a Hilbert space
\hilbspace\ are the vectors of unit norm, conventionally denoted by a
\emph{ket} $\ket{y} \in \hilbspace$. These are equivalently given by
morphisms $\mathbb{C} \to \hilbspace$ in $\cat{FHilb}$ that map
$z \in \mathbb{C}$ to $z\ket{y} \in \hilbspace$.  Dually, the
functional $\hilbspace \to \mathbb{C}$ which maps $y \in \hilbspace$
to the inner product $\braket{x}{y}$ is conventionally written as a
\emph{bra} $\bra{x}$. Morphisms $\hilbspace \to \mathbb{C}$ are also
called \emph{effects}.

In fact, \cat{FHilb} is a dagger rig category. The \emph{dagger} of
linear map $f \colon A \to B$ is uniquely determined via the inner
product by $\braket{f(x)}{y}=\braket{x}{f^\dag(y)}$.  The dagger of a
state is an effect, and vice versa.  In quantum computing, pure states
evolve along unitary gates. These are exactly the morphisms that are
unitary in the sense of dagger categories in that
$f^\dag \circ f = \id$ and $f \circ f^\dag = \id$, exhibiting the
groupoid $\cat{Unitary}$ as a dagger subcategory of $\cat{FHilb}$.

There is a way to translate the category $\cat{FPInj}$ of finite sets
and partial injections to the category $\cat{FHilb}$, that sends
$\{0,\ldots,n-1\}$ to $\mathbb{C}^n$. This translation preserves
composition, identities, tensor product, direct sum, and dagger: it is
a dagger rig functor $\ell^2 \colon \cat{FPInj} \to \cat{FHilb}$, that
restricts to a dagger rig functor $\cat{FinBij} \to \cat{Unitary}$~\cite{heunen:ltwoc}.
Thus reversible computing ($\cat{FinBij}$) is to classical reversible
theory ($\cat{FPInj}$) as quantum computing ($\cat{Unitary}$) is to
quantum theory ($\cat{FHilb}$).  In particular, in this way, the
Boolean controlled-controlled-not function (known as the \emph{Toffoli
  gate}), which is universal for reversible computing, transfers to a
quantum gate with the same name that acts on vectors.

\subsection{The Hadamard Mystery} 
\label{sub:quantumeff}

Shi established that quantum computing can be characterized as a
relatively small increment over classical
computing~\cite{10.5555/2011508.2011515}. The precise statement below
is adapted from Aharonov's reformulation of Shi's result~\cite{aharonov:toffolihadamard}.

\begin{theorem}
  The set consisting of just the Toffoli and Hadamard gates is
  computationally universal for quantum computing. By
  \emph{computationally universal}, we mean that the set can simulate,
  to within $\epsilon$-error, an arbitrary quantum circuit of $n$
  qubits and $t$ gates with only poly-logarithmic overhead in
  $(n,t,1/\epsilon)$.
\end{theorem}

\noindent The result may appear counter-intuitive since it omits any
reference to complex numbers. The subtlety is that \emph{computational
  universality} allows arbitrary---but efficient---encodings of
complex vectors and matrices.

The significance of this result is the following. The Toffoli gate is
known to be universal for classical computing over finite
domains~\cite{10.1007/3-540-10003-2104}. Thus, in one sense, a quantum
computation is nothing but a classical computation that is given
access to one extra primitive, the Hadamard transform.

Once expressed in rig categories, this result allows novel
characterizations of quantum computing. The key to these
characterizations is that quantum gates are \emph{not} black boxes in
rig categories: they are ``white boxes'' constructed from $\oplus$ and
other primitives which means that can be decomposed and recomposed
during rewriting using the coherence conditions of rig categories. For
example, while a circuit theory will allow one to derive that
$TT = S$, it is unable to provide justification for this in terms of
the definitions of $S$ and $T$. On the other hand, the rig model
reduces this equation to the bifunctoriality of $\oplus$ and the
definitions of $S$ and $T$. This style of reasoning enabled two recent
characterizations of quantum computing by (universal) categorical
constructions. In the first paper~\cite{quantum-eff}, Hadamard is
recovered by two copies of $\PiLang$ mediated by one equation for
\emph{complementarity}. In the second
paper~\cite{caretteetal2024sqrtpi}, Hadamard is recovered by
postulating the existence of square roots for certain morphisms, i.e.
the existence of morphisms $\sqrt{f}$ such that
$\sqrt{f} \circ \sqrt{f} = f$.

\subsection{Quantum Information Effects}\label{subsec:qie}

Information effects~\cite{jamessabry:infeff,heunenkaarsgaard:qie}
emulate the dynamics of open systems using the reversible dynamics of
closed systems, by extending the latter with the ability to hide parts
of the input and output spaces. This allows auxiliary states to be
prepared and (parts of) the output to be discarded -- as a
consequence, measurement is recovered in the quantum case.

This idea comes from the theory of quantum computation, where
\emph{Stinespring's dilation theorem}~\cite{stinespring:positive} (see
also \cite{huotstaton:universal,huotstaton:channels,%
  heunenkaarsgaard:bennettstinespring}) provides a recipe for
reconstructing the reversible dynamics of an irreversible quantum
channel by outfitting it with an auxiliary system that can be used as
a sink for the data that the process discards.

Concretely, given a quantum channel $\Lambda : H \to K$ (which can be
thought of as a quantum circuit where measurements can occur),
Stinespring's theorem argues that it is always possible to factor a
quantum channel as an isometry (a kind of injective quantum map) $V :
H \to H \otimes G$ followed by a projection $\pi_1 : H \otimes G \to
H$. Note that projection is not as innocuous as it is in the classical
case, since it may lead to the formation of probabilistic (mixed)
states. In turn, it can be shown that every isometry can be realized
by fixing a part of the input to a reversible (unitary) quantum
map~\cite{huotstaton:channels}: in other words, every isometry factors
as an injection $\iota_1 : H \to H \oplus E$ (fixing a part of the
inputs) followed by a reversible (unitary) quantum map $U : H \oplus E
\to K$.

Putting these two factorizations together, we get that any quantum
channel (i.e., irreversible quantum process) $\Lambda : H \to K$ factors
(in an essentially unique way) into three stages:
\begin{equation*}
  H \xrightarrow{\text{prepare auxiliary state}} H \oplus G
  \xrightarrow{\text{reversible dynamics}} K \otimes E
  \xrightarrow{\text{discard environment}} K
\end{equation*}
Put another way, every quantum channel can be written as a unitary in
which a part of the input (corresponding to the subsystem $G$ above
receiving a fixed state) and a part of the output (corresponding to the
subsystem $E$ above that is discarded after use) is hidden from view.

This factorization is clearly reminiscent of the fundamental theorem
of reversible computing and the LR-construction of
Sec.~\ref{sec:fundamental} . More concretely, we define the hiding of
the input and output through a stack of two separate effects, which
turn out to correspond to \emph{Arrows}, and so give suitable notions
of effectful programs with sequential and parallel composition. To
that end, we define $\PiLang$ with allocation to have the same base
types $b$ as $\PiLang$, and with the combinator type
$b_1 \rightarrowtail b_2$. Terms in $\PiLang$ with allocation are
given by the formation rule:
$$
\frac{u \of b_1 + b_3 \fromto b_2}{\mathit{lift}(u) \of b_1 \rightarrowtail b_2}
$$
That is, terms in $\PiLang$ with allocation are given by $\PiLang$
terms where part of their input is hidden.  Additionally, we consider two
terms in $\PiLang$ with allocation to be equal if they are equal up to
an arbitrary term applied on the hidden part alone. These terms can be
composed in sequence by:
$$
\mathit{lift}(u) \ggg \mathit{lift}(v) =
\mathit{lift}(\associp \seqq (u \oplus \mathit{id}) \seqq v)
$$
and it can be shown that this is associative, and that
$\mathit{lift}(\unitep)$ acts as the identity with respect to
composition. More generally, every $\PiLang$ term $u : b_1 \to b_2$
can be turned into one $\mathit{arr}(u) \of b_1 \rightarrowtail b_2$
that acts as $u$ does by letting it hide only the empty system, that
is:
$$\mathit{arr}(u) = \mathit{lift}(\unitep \seqq u)$$
Finally, it can be shown that this also allows a parallel composition
$\mathit{lift}(u) \ppp \mathit{lift}(v)$ to be defined, giving it all
the structure of an Arrow.

A consequence of these definitions is that can define a new term
$$\mathit{alloc} = \mathit{lift}(\unitepl) \of 0 \rightarrowtail b$$
which can be thought of as allocating a constant value from a hidden
heap. It can be shown that this ability to allocate new constants is
enough to extend $\PiLang$ with the ability to perform arbitrary
injections $\mathit{inl} \of b_1 \rightarrowtail b_1 + b_2$ and
$\mathit{inr} \of b_2 \rightarrowtail b_1 + b_2$, and to do classical
cloning via a term $\mathit{clone} : b \rightarrowtail b \times b$.
This is the first step of two in recovering open system dynamics from
their reversible foundations.

The second step is a study in duality: to further extend $\PiLang$
with allocation and hiding, we introduce yet another arrow whose base
types are the same as those of $\PiLang$, and whose combinator types
are given by a new type $b_1 \rightsquigarrow b_2$. Terms in this new
layer are formed by the rule:
$$
\frac{v \of b_1 \rightarrowtail b_2 \times b_3}
{\mathit{lift}(v) \of b_1 \rightsquigarrow b_2}
$$
and, by analogy to the previous definitions, one can define sequential
and parallel composition, identities, and the lifting of arbitrary
terms $v : b_1 \rightarrowtail b_2$ to $\mathit{arr}(v) : b_1
\rightsquigarrow b_2$ by adjoining the trivial system $1$. This gives it
the structure of an arrow. A consequence of this is that arbitrary
data can now be discarded via a term $\mathit{discard} \of b
\rightsquigarrow 1$, and by combining this with parallel composition
and the unitor $\unitet \of b \times 1 \rightarrowtail b$, we obtain
projections $\mathit{fst} \of b_1 \times b_2 \rightsquigarrow b_1$ and
$\mathit{snd} \of b_1 \times b_2 \rightsquigarrow b_2$, completing
our journey from fully reversible to fully irreversible dynamics.

While it seems clear that we can recover irreversible classical
computing from their reversible foundations by extending them with the
ability to allocate constants and hide arbitrary data, it is less
clear that one can also recover irreversible quantum computing this
way. Surprisingly, this is so, with measurement (i.e., the map that
sends a quantum state to its post measurement state after measurement
in the computational basis) given the (classically nonsensical)
term:
$$
\mathit{measure} = \mathit{clone} \ggg \mathit{fst} \of b
\rightsquigarrow b \enspace.
$$
and it can be verified that this recovers the usual Born rule
assigning probabilities to classical measurement outcomes.

\section{Conclusions and Future Research} 

Pragmatically, reversible computing, reversible programming languages,
and bidirectional methods in computing have unified many of the
original software engineering instances of reversibility which is
clearly a positive contribution to the field of computer science. 

But it has been 32 years since Baker stated that a ``physics
revolution is brewing in computer science'' and it is fair to ask to
what extent has this ``revolution'' been realized?

From the very beginning, one of the most common arguments for the
physics revolution in computer science has been the potential to
drastically reduce the energy needs of computation. The reasoning is
that only irreversible operations need to dissipate heat and hence
reversible computing can in principle operate near the thermodynamic
limit. Despite its theoretical plausibility and its experimental
validation~\cite{5392446,Berut2012}, the promise of drastically more
energy-efficient computers has not yet materialized.

We argue that the real revolution is more of a conceptual one,
affecting what we mean by computation, logic, and information, and
unifying them in ways that give new insights about the nature of logic
and the fundamental limits of information processing by computers.

On the one hand, treating information as a first-class entity promotes
several ad hoc techniques to the fold of well-established logical and
semantic techniques. Examples includes the methods used in
applications such as quantitative information-flow
security~\cite{1159651}, differential
privacy~\cite{10.1007/117870061}, energy-aware
computing~\cite{10.1145/1324177.1324180,10.1145/605397.605411}, VLSI
design~\cite{497613}, and biochemical models of
computation~\cite{10.1007/978-3-540-85101-16}.

On the other hand, reversible computing is the first key to
understanding how Nature computes, how to integrate computational
models with their physical environments, and to explore new modes of
computation such as molecular computing, biologically-inspired
computing, neuromorphic computing, emerging phenomena in complex
systems, and of course quantum computing.

\begin{credits}

\subsubsection{\ackname}
This material is based upon work supported by the National. Science
Foundation under Grant No. 1936353.

\subsubsection{\discintname}
The authors have no competing interests to declare that are
relevant to the content of this article. 
\end{credits}

%

\bibliographystyle{splncs04}
\bibliography{bibliography}

\begin{thebibliography}{10}
\providecommand{\url}[1]{\texttt{#1}}
\providecommand{\urlprefix}{URL }
\providecommand{\doi}[1]{https://doi.org/#1}

\bibitem{abramsky:reversible}
Abramsky, S.: A structural approach to reversible computation. Theoretical
  Computer Science  \textbf{347},  441--464 (2005)

\bibitem{aharonov:toffolihadamard}
Aharonov, D.: A simple proof that {T}offoli and {H}adamard are quantum
  universal (2003),
  \href{https://arxiv.org/abs/quant-ph/0301040}{arXiv:quant-ph/0301040}

\bibitem{andresmartinezheunenkaarsgaard:universal}
Andrés-Martínez, P., Heunen, C., Kaarsgaard, R.: Universal properties of
  partial quantum maps (2022),
  \href{https://arxiv.org/abs/2206.04814}{arXiv:2206.04814}

\bibitem{AubertRCC23}
Aubert, C.: The correctness of concurrencies in (reversible) concurrent
  calculi. Journal of Logical and Algebraic Methods in Programming p. 100924
  (2023). \doi{10.1016/j.jlamp.2023.100924}

\bibitem{10.1007/978-3-642-19861-8_9}
Axelsen, H.B.: Clean translation of an imperative reversible programming
  language. In: Knoop, J. (ed.) Compiler Construction. pp. 144--163. Springer
  Berlin Heidelberg, Berlin, Heidelberg (2011)

\bibitem{HBA}
Axelsen, H.B.: Private communication (2015)

\bibitem{baez-dolan}
Baez, J.C., Dolan, J.: From finite sets to {F}eynman diagrams (2000),
  \href{https://arxiv.org/abs/math/0004133}{arXiv:000413}

\bibitem{10.5555/645648.664816}
Baker, H.G.: {NREVERSAL of Fortune -- The Thermodynamics of Garbage
  Collection}. In: Proceedings of the International Workshop on Memory
  Management. pp. 507--524. IWMM '92, Springer-Verlag, Berlin, Heidelberg
  (1992)

\bibitem{5391327}
Bennett, C.H.: Logical reversibility of computation. IBM Journal of Research
  and Development  \textbf{17}(6),  525--532 (1973). \doi{10.1147/rd.176.0525}

\bibitem{Berut2012}
B{\'e}rut, A., Arakelyan, A., Petrosyan, A., Ciliberto, S., Dillenschneider,
  R., Lutz, E.: Experimental verification of landauer's principle linking
  information and thermodynamics. Nature  \textbf{483}(7388),  187--189 (Mar
  2012). \doi{10.1038/nature10872}, \url{https://doi.org/10.1038/nature10872}

\bibitem{10.1007/978-3-540-85101-16}
Cardelli, L., Zavattaro, G.: On the computational power of biochemistry. In:
  Horimoto, K., Regensburger, G., Rosenkranz, M., Yoshida, H. (eds.) Algebraic
  Biology. pp. 65--80. Springer Berlin Heidelberg, Berlin, Heidelberg (2008)

\bibitem{caretteetal2024sqrtpi}
Carette, J., Heunen, C., Kaarsgaard, R., Sabry, A.: With a few square roots,
  quantum computing is as easy as pi. Proceedings of the ACM on Programming
  Languages  \textbf{8}(POPL),  546--574 (2024)

\bibitem{CARETTE202215}
Carette, J., James, R.P., Sabry, A.: Embracing the laws of physics: Three
  reversible models of computation. Advances in Computers, vol.~126, pp.
  15--63. Elsevier (2022). \doi{https://doi.org/10.1016/bs.adcom.2021.11.009},
  \url{https://www.sciencedirect.com/science/article/pii/S0065245821000838}

\bibitem{10.1007/978-3-662-49498-1}
Carette, J., Sabry, A.: Computing with semirings and weak rig groupoids. In:
  Thiemann, P. (ed.) Programming Languages and Systems. pp. 123--148. Springer
  Berlin Heidelberg, Berlin, Heidelberg (2016)

\bibitem{quantum-eff}
Carette, J., Heunen, C., Kaarsgaard, R., Sabry, A.: The quantum effect: A
  recipe for {QuantumPi} (2023),
  \href{https://arxiv.org/abs/2302.01885}{arXiv:2302.01885}

\bibitem{10.1145/3498667}
Choudhury, V., Karwowski, J., Sabry, A.: Symmetries in reversible programming:
  From symmetric rig groupoids to reversible programming languages. Proc. ACM
  Program. Lang.  \textbf{6}(POPL) (jan 2022). \doi{10.1145/3498667},
  \url{https://doi.org/10.1145/3498667}

\bibitem{15897363-af72-3dac-82e6-fde144ad66c0}
Church, A.: A set of postulates for the foundation of logic. Annals of
  Mathematics  \textbf{33}(2),  346--366 (1932),
  \url{http://www.jstor.org/stable/1968337}

\bibitem{10.1007/978-3-540-28644-8_19}
Danos, V., Krivine, J.: Reversible communicating systems. In: Gardner, P.,
  Yoshida, N. (eds.) CONCUR 2004 - Concurrency Theory. pp. 292--307. Springer
  Berlin Heidelberg, Berlin, Heidelberg (2004)

\bibitem{10.1007/117870061}
Dwork, C.: Differential privacy. In: Bugliesi, M., Preneel, B., Sassone, V.,
  Wegener, I. (eds.) Automata, Languages and Programming. pp. 1--12. Springer
  Berlin Heidelberg, Berlin, Heidelberg (2006)

\bibitem{fredkin-conservative-1982}
Fredkin, E., Toffoli, T.: Conservative logic. International Journal of
  Theoretical Physics  \textbf{21}(3),  219--253 (Apr 1982).
  \doi{10.1007/BF01857727}, \url{https://doi.org/10.1007/BF01857727}

\bibitem{10.1007/978-3-540-24754-8_21}
Gl{\"u}ck, R., Kawabe, M.: Derivation of deterministic inverse programs based
  on {LR} parsing. In: Kameyama, Y., Stuckey, P.J. (eds.) Functional and Logic
  Programming. pp. 291--306. Springer Berlin Heidelberg, Berlin, Heidelberg
  (2004)

\bibitem{GLUCK2023113429}
Glück, R., Yokoyama, T.: Reversible computing from a programming language
  perspective. Theoretical Computer Science  \textbf{953},  113429 (2023).
  \doi{https://doi.org/10.1016/j.tcs.2022.06.010},
  \url{https://www.sciencedirect.com/science/article/pii/S0304397522003619}

\bibitem{heunen:ltwo}
Heunen, C.: On the functor $\ell^2$. In: Computation, Logic, Games, and Quantum
  Foundations. pp. 107--121. Springer (2013)

\bibitem{heunenkaarsgaard:bennettstinespring}
Heunen, C., Kaarsgaard, R.: Bennett and stinespring, together at last. In:
  Proceedings 18th International Conference on Quantum Physics and Logic (QPL
  2021). Electronic Proceedings in Theoretical Computer Science, vol.~343, pp.
  102--118. OPA (2021). \doi{10.4204/EPTCS.343.5}

\bibitem{heunenkaarsgaard:qie}
Heunen, C., Kaarsgaard, R.: Quantum information effects. Proceedings of the ACM
  on Programming Languages  \textbf{6}(POPL),  1--27 (2022)

\bibitem{heunenkaarsgaardkarvonen:arrows}
Heunen, C., Kaarsgaard, R., Karvonen, M.: Reversible effects as inverse arrows.
  In: Proceedings of the Thirty-Fourth Conference on the Mathematical
  Foundations of Programming Semantics (MFPS XXXIV). Electronic Notes in
  Theoretical Computer Science, vol.~341, pp. 179--199. Elsevier (2018)

\bibitem{heunenkarvonen:daggermonads}
Heunen, C., Karvonen, M.: Monads on dagger categories. Theory and Applications
  of Categories  \textbf{31},  1016--1043 (2016)

\bibitem{heunenvicary:cqt}
Heunen, C., Vicary, J.: Categories for quantum theory. Oxford University Press
  (2019)

\bibitem{10.1145/359576.359585}
Hoare, C.A.R.: Communicating sequential processes. Commun. ACM  \textbf{21}(8),
   666–677 (aug 1978). \doi{10.1145/359576.359585},
  \url{https://doi.org/10.1145/359576.359585}

\bibitem{hughes:arrows}
Hughes, J.: Programming with {A}rrows. In: Advanced Functional Programming.
  Lecture Notes in Computer Science, vol.~3622, pp. 73--129. Springer (2005).
  \doi{10.1007/11546382_2}

\bibitem{huotstaton:channels}
Huot, Staton, S.: Quantum channels as a categorical completion. In: Proceedings
  of the ACM/IEEE Symposium on Logic in Computer Science. vol.~35, pp. 1--13
  (2019)

\bibitem{huotstaton:universal}
Huot, M., Staton, S.: Universal properties in quantum theory. In: Selinger, P.,
  Chiribella, G. (eds.) Proceedings of the 15th International Conference on
  Quantum Physics and Logic (QPL 2018). Electronic Proceedings in Theoretical
  Computer Science, vol.~287, pp. 213--224. Open Publishing Association (2018).
  \doi{10.4204/EPTCS.287.12}

\bibitem{jacobsheunenhasuo:arrows}
Jacobs, B., Heunen, C., Hasuo, I.: Categorical semantics for {A}rrows. Journal
  of Functional Programming  \textbf{19}(3--4),  403--438 (2009).
  \doi{10.1017/S0956796809007308}

\bibitem{jamessabry:infeff}
James, R.P., Sabry, A.: Information effects. In: POPL '12: Proceedings of the
  39th Annual ACM SIGPLAN-SIGACT Symposium on Principles of programming
  languages. pp. 73--84. ACM (2012). \doi{10.1145/2103656.2103667}

\bibitem{kaarsgaardglueck:rfcl}
Kaarsgaard, R., Gl{\"u}ck, R.: A categorical foundation for structured
  reversible flowchart languages: {S}oundness and adequacy. Logical Methods in
  Computer Science  \textbf{14}(3) (2018)

\bibitem{RKthesis}
Kaarsgaard, R.: The Logic of Reversible Computing: Theory and Practice. Ph.D.
  thesis, Department of Computer Science, University of Copenhagen (2018)

\bibitem{kastl:inverse}
Kastl, J.: Algebraische {M}odelle, {K}ategorien und {G}ruppoide, Studien zur
  {A}lgebra und ihre {A}nwendungen, vol.~7, chap. Inverse categories, pp.
  51--60. Akademie-{V}erlag {B}erlin (1979)

\bibitem{10.1007/BFb0063106}
Kelly, G.M.: Coherence theorems for lax algebras and for distributive laws. In:
  Kelly, G.M. (ed.) Category Seminar. pp. 281--375. Springer Berlin Heidelberg,
  Berlin, Heidelberg (1974)

\bibitem{5392446}
Landauer, R.: Irreversibility and heat generation in the computing process. IBM
  Journal of Research and Development  \textbf{5}(3),  183--191 (1961).
  \doi{10.1147/rd.53.0183}

\bibitem{laplaza:distributivity}
Laplaza, M.L.: Coherence for distributivity. In: Coherence in categories. pp.
  29--65. No.~281 in Lecture Notes in Mathematics, Springer (1972)

\bibitem{lecerf1963machines}
Lecerf, Y.: Machines de {T}uring reversibles. Comptes Rendus hebdomadaires des
  seances de l'academie des sciences  \textbf{257},  2597--2600 (1963)

\bibitem{leinster:categorytheory}
Leinster, T.: Basic category theory. Cambridge University Press (2014)

\bibitem{10.1145/1324177.1324180}
Ma, X., Huang, J., Lombardi, F.: A model for computing and energy dissipation
  of molecular qca devices and circuits. J. Emerg. Technol. Comput. Syst.
  \textbf{3}(4) (jan 2008). \doi{10.1145/1324177.1324180},
  \url{https://doi.org/10.1145/1324177.1324180}

\bibitem{maclane:associativity}
{Mac Lane}, S.: Natural associativity and commutativity. Rice University
  Studies  \textbf{49}(4) (1963)

\bibitem{497613}
Macii, E., Poncino, M.: Exact computation of the entropy of a logic circuit.
  In: Proceedings of the Sixth Great Lakes Symposium on VLSI. pp. 162--167
  (1996). \doi{10.1109/GLSV.1996.497613}

\bibitem{McC56}
McCarthy, J.: The inversion of functions defined by turing machines. In:
  C.E.~Shannon, J.M. (ed.) Automata Studies, Annals of Mathematical Studies,
  pp. 177--181. No.~34, Princeton University Press (1956)

\bibitem{DBLP:books/sp/Milner80}
Milner, R.: A Calculus of Communicating Systems, Lecture Notes in Computer
  Science, vol.~92. Springer (1980). \doi{10.1007/3-540-10235-3},
  \url{https://doi.org/10.1007/3-540-10235-3}

\bibitem{Milner_1992}
Milner, R.: Functions as processes. Mathematical Structures in Computer Science
   \textbf{2}(2),  119–141 (1992). \doi{10.1017/S0960129500001407}

\bibitem{moggi}
Moggi, E.: Notions of computations and monads. Information and Computation
  \textbf{93},  55--92 (1991)

\bibitem{nielsenchuang:quantum}
Nielsen, M.A., Chuang, I.: Quantum Computation and Quantum Information.
  Cambridge University Press (2002)

\bibitem{10.1007/11690634_17}
Phillips, I., Ulidowski, I.: Reversing algebraic process calculi. In: Aceto,
  L., Ing{\'o}lfsd{\'o}ttir, A. (eds.) Foundations of Software Science and
  Computation Structures. pp. 246--260. Springer Berlin Heidelberg, Berlin,
  Heidelberg (2006)

\bibitem{1159651}
Sabelfeld, A., Myers, A.: Language-based information-flow security. IEEE
  Journal on Selected Areas in Communications  \textbf{21}(1),  5--19 (2003).
  \doi{10.1109/JSAC.2002.806121}

\bibitem{10.5555/2011508.2011515}
Shi, Y.: Both {Toffoli} and {Controlled-NOT} need little help to do universal
  quantum computing. Quantum Info. Comput.  \textbf{3}(1),  84–92 (jan 2003)

\bibitem{stinespring:positive}
Stinespring, W.F.: Positive functions on {C*}-algebras. Proceedings of the
  American Mathematical Society  \textbf{6}(2),  211--216 (1955).
  \doi{10.2307/2032342}

\bibitem{10.1007/978-3-319-20860-2_13}
Thomsen, M.K., Kaarsgaard, R., Soeken, M.: Ricercar: A language for describing
  and rewriting reversible circuits with ancillae and its permutation
  semantics. In: Krivine, J., Stefani, J.B. (eds.) Reversible Computation. pp.
  200--215. Springer International Publishing, Cham (2015)

\bibitem{10.1007/3-540-10003-2104}
Toffoli, T.: Reversible computing. In: {de Bakker}, J., {van Leeuwen}, J.
  (eds.) Automata, Languages and Programming. pp. 632--644. Springer Berlin
  Heidelberg, Berlin, Heidelberg (1980)

\bibitem{doi:10.1137/S0097539703432165}
van Tonder, A.: A lambda calculus for quantum computation. SIAM Journal on
  Computing  \textbf{33}(5),  1109--1135 (2004).
  \doi{10.1137/S0097539703432165},
  \url{https://doi.org/10.1137/S0097539703432165}

\bibitem{doi.org/10.1112/plms/s2-42.1.230}
Turing, A.M.: On computable numbers, with an application to the
  entscheidungsproblem. Proceedings of the London Mathematical Society
  \textbf{s2-42}(1),  230--265 (1937).
  \doi{https://doi.org/10.1112/plms/s2-42.1.230},
  \url{https://londmathsoc.onlinelibrary.wiley.com/doi/abs/10.1112/plms/s2-42.1.230}

\bibitem{yanofskymannucci:quantumcomputing}
Yanofsky, N., Mannucci, M.A.: Quantum Computing for Computer Scientists.
  Cambridge University Press (2008)

\bibitem{10.1145/362342.362360}
Zelkowitz, M.V.: Reversible execution. Commun. ACM  \textbf{16}(9), ~566 (sep
  1973). \doi{10.1145/362342.362360},
  \url{https://doi.org/10.1145/362342.362360}

\bibitem{10.1145/605397.605411}
Zeng, H., Ellis, C.S., Lebeck, A.R., Vahdat, A.: Ecosystem: managing energy as
  a first class operating system resource. In: Proceedings of the 10th
  International Conference on Architectural Support for Programming Languages
  and Operating Systems. p. 123–132. ASPLOS X, Association for Computing
  Machinery, New York, NY, USA (2002). \doi{10.1145/605397.605411},
  \url{https://doi.org/10.1145/605397.605411}

\end{thebibliography}
\end{document}